\documentclass[12pt]{article}
\usepackage{mathtext}
\usepackage{amsfonts}
\usepackage{amssymb}
\usepackage{amsmath}
\usepackage{graphicx}
\usepackage{courier}
\usepackage{subfig}
\renewcommand{\d}{\delta}
\renewcommand{\b}{\beta}
\newcommand{\g}{\gamma}

\newcommand{\ar}{\longrightarrow}
\newcommand{\w}{\omega}

\newcommand{\la}{\lambda}
\renewcommand{\a}{\alpha}
\begin{document}
\title{Entanglement in systems of oscillators and quantum computations}
\date{}
\author{Y.I.Ozhigov\thanks{This article was supported by Russian
Fond for Fundamental Researches, grant 12-01-00475-a.}\\
\\
{\it Moscow State University of M.V.Lomonosov}\\
{\it Institute of Physics and Technology RAS}}
\maketitle

\begin{abstract}
It is shown that quantum devices based only on oscillators cannot serve as the universal quantum computer, despite of entanglement in such devices, which we roughly estimate for the ideal case and for the harmful entanglement with photonic modes. We show that quasi-particles are the native shell for the entanglement already for ground state, in contast to the free electromagnetic field where vacuum state does not produce entanglement at all. 
\end{abstract}

\section{Introduction}

Entanglement of a wave function $\Psi (r_1,r_2,\ldots,r_n)$ of $n$ particles means that it cannot be represented as the product $\Psi_1(r_1)\Psi_2(r_2)\ldots\Psi_n(r_n)$ of one particle wave functions. The analogous definition takes place for the entanglement between two subsets of $\{ 1,2,\ldots,n\}$ that is the so-called bi-partite entanglement. This fundamental notion plays the key role in quantum information processing, especially in quantum computing, where entanglement is the necessary property of quantum states, which arise in the course of quantum computations, if it pretends to be faster than its classical counterpart. Entanglement is the physical phenomenon, which results in quantum non locality, that is the violation of Bell inequality, detected in numerous experiments. 

The important role, which entanglement plays in information processing explains that the most elaborated numerical measures of this property used the discrete form of quantum state representation, e.g. its qubit form
$$
|\Psi\rangle = \sum\limits_{j=0}^{N-1}\la_j|j\rangle
$$
where $N=2^n$, $n$ is the number of qubits. In this representation the product of wave functions turns to be tensor product, an we can pass to the continuous representation and vice versa with some lack in the accuracy that follows from the approximate equation
$$
\Psi (\bar R)=\sum\limits_J\Psi (\bar R^J)\d_J
$$
where $\d_j$ is proportional to the characteristic function of $j$-th cube in the configuration space for the continuous wave function, which plays the role of $|j\rangle$. Because of exponentially huge dimentionality of Hilbert space (values of $J$) the last equation cannot be accurate even for relatively small $n$, and just this non accuracy dramatically complicates the application of discrete entanglement measures to the continuous case. 

The simple and robust measure of entanglement for many qubits (and qu-dits - $d$-levels quantum elements) has been proposed in the paper \cite{Ch}. It is the minimal value $Ch(\Psi )$ of Shannon entropy of quantum state amplitude distribution: $-\sum_j|\la_j|ln(|\la_j|)$ taken after all possible one-qubit unitary operations on the given quantum state $\Psi$. Entanglement measure $Ch$ possesses all properties of entropy, including the different form of additivity, and has extremal value on $W$- and $GHZ$- states; there are efficient algorithms for its computation and it can be generalized on fermionic quantum states. But this measure hardly can be applied to states of continuous objects, like the field, because of its substantially discrete character. 

Instead of it we use the simple idea (look at \cite{Ak}) how to check entanglement for a function with two argument by the sequential differentiation. The function $f(x,y)$ is non entangled if and only if
$$
\frac{\partial^2ln\ f}{\partial x\partial y}=0.
$$
This will be the indicator of bi-partite entanglement if we sum up this values over all $x\in X,\ y\in Y$ for the division $X,Y$ of our configuration space. Following Born rule, we obtain the bi-partite measure of entanglement for the continuous wave function $\Psi (u_1,u_2,\ldots,u_n)$ in the form
\begin{equation}
E_{i,j}=\int\limits_{{\cal K}}|\Psi |^2\left|\frac{\partial^2ln\ \Psi}{\partial u_i\partial u_j}\right|du_1du_2\ldots du_n
\label{ent_measure}
\end{equation}

The measure of bi-partite entanglement given by (\ref{ent_measure}) has the transarent sense in the case of kernel-type wave functions of systems with quadratic Lagranjians, which have the form $\Psi =F(t)exp(\frac{iS}{h})$ where $S$ is the action along the classical path. For example, for the interaction between oscillating charged particles and the field, represented by the amplitude $a$ the action has the form  $S=S_{particles}+S_{field}-\frac{1}{c}ja$. For the small $t$ the value of our measure will be independent of the coordinates: $E_{particles,\ field}=\frac{e}{hc}$. For the general wave functions it is not true, because the summing of such values on the different paths changes entanglement. 
We will use this measure for the entanglement of the two kind of media: interacting harmonic oscillators and electromagnetic field. 

However, in some cases it is more convenient to use the canonic expression of the bi-partite entanglement measure, given by the formula
\begin{equation}
{\cal M}_{ij}=-tr(\rho_i\ ln(\rho_i))=-tr(\rho_j\ ln(\rho_j)).
\label{canonic_ent}
\end{equation}

\section{Interacting oscillators}

A system of interacting oscillators form the basis for the most elaborated technology of quantum computer: excited levels of ions in Paul trap. There quantum gates on excited ($|1\rangle$) and ground state ($|0\rangle$) levels of ions are organized with the ancillary qubits, which role plays phonons of the mechanical oscillations of ions in trap (see \cite{Zo},\cite{Bu}). For the entangling of ion levels with the oscillation mode classical laser field is used. This model: real charged oscillators plus field form the abstract model of such a quantum computer. The field can be represented as the ensemble of (imaginary) oscillators as well; we factually will take up entanglement in oscillator systems. We start with the one dimensional system of real charged particles of the same mass, each of which has its equilibrium position, and let $u_j$ denote the shift from the equilibrium for $j$-th particle. If $\kappa$ is Gook coefficient for the force between the neigbor particles, Hamiltonian of the whole ensemble has the form
\begin{equation}
H=\sum\limits_{n=1}N(\frac{p_n^2}{2m}+\kappa u_n^2)-\kappa\sum\limits_{n=1}^Nu_nu_{n+1}.
\label{stringHam}
\end{equation}
Let the total number $N$ of oscillators be so large that we can ignore boundary conditions (alternatively, we can enclose the string of oscillators to the ring without bounds). When passing to a media (field) we launch $N$ to infinity and use integrals instead of sums.

Even for not large $N$ it is impossible to operate with this Hamiltonian directly because of the summand of interaction $u_nu_{n+1}$, which cause entanglement. We must use the canonic transformation to get rid of interaction:

\begin{equation}
\begin{array}{ll}
&u_n=\frac{1}{\sqrt N}\sum\limits_qU_qe^{-iqnd},\\
&U_q=\frac{1}{\sqrt N}\sum\limits_nu_ne^{iqnd}.
\end{array}
\label{can}
\end{equation}
where $d=2\pi/N$.

This is the canonic transformation, which elimitanes the interaction; in new coordinates $U$ we have the system of non interacting oscillators. 

Using the definition of impulse $p_n=\frac{h}{i}\frac{\partial}{\partial u_n}$ and rules of differentiation, we obtain formulas for the transformation of impulses:

\begin{equation}
\begin{array}{ll}
&p_n=\frac{1}{\sqrt N}\sum\limits_qP_qe^{iqnd},\\
&P_q=\frac{1}{\sqrt N}\sum\limits_np_ne^{-iqnd},
\label{canp}
\end{array}
\end{equation}

where $P_n=\frac{h}{i}\frac{\partial}{\partial U_n}$.

\begin{figure}
\centering
\caption{Quasi-particles}
\vspace{150mm}
\makebox[380mm][l]{\includegraphics{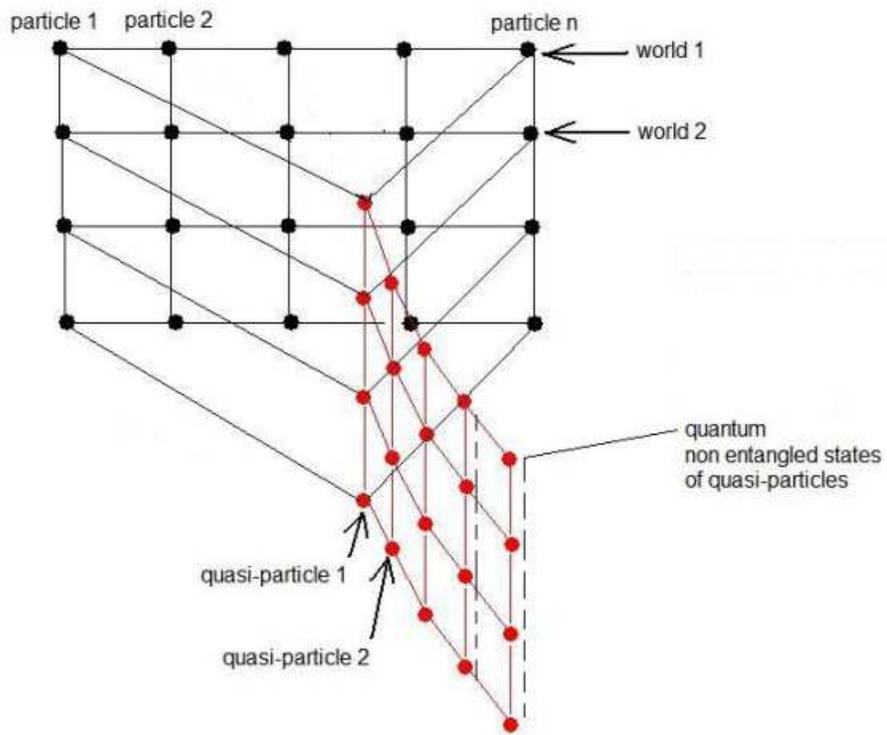}}%
\end{figure}

Canonic transformation is linear and transforms any small cube of the division of the configuration space to the cube due the orthogonality of (\ref{can}). We thus can treat the canonic transform as the permutation of basic vectors in Hilbert space of quantum states, like $CNOT$ two-qubit operation: $|x,y\rangle\ar |x,x+y\ (mod\ 2)\rangle$. Canonic trasformation is one-to-one corresponding between points in the configuration space for our system of oscillators (see Figure 1). This transformation is entangling and we will estimate the measure of entanglement it produces. It is important, that this transformation eliminate the initial entanglement, which means that it incapsulate this entanglement into quasi-particles, called phonons. Spectrum of phonons is discrete (see (\ref{frequences})); we will see that the discrete form of spectrum is typical for ensembles of oscillator type with entanglement.

Turn the coordinate system for $q$ so that this parameter, replacing $n$, takes value from the symmetric segment. Instead of $q+q'=N$ we then write $q+q'=0$. The relation of inequality is induced form the old set $1,2,\ldots,N$ so that pairs where $q>-q$ are approximately the half.

Rewriting Hamiltonian in the new coordinates, we have:
$$
\begin{array}{ll}
H=&\sum\limits_{n=1}^N\frac{1}{2mN}(\sum\limits_{q,q'}P_qP_{q'}e^{1(q+q')nd})+\frac{K}{N}\sum\limits_qU_qU_{q'}\\
\ &-\frac{K}{N}\sum\limits_{q,q'}(U_qU_{q'}e^{-iqnd}e^{-iq'(n+1)d})=\\
&=\frac{1}{2mN}\sum\limits_qP_qP_{-q}-\frac{K}{N}\sum\limits_{q,q'}U_qU_{q'}(\sum
\limits_{n=1}^Ne^{-ind(q+q')})e^{-iq'd}+\frac{K}{N}\sum\limits_qU_qU_{q'}=\\
&=\frac{1}{2mN}\sum\limits_qP_qP_{-q}-\frac{K}{N}\sum\limits_{q}U_qU_{-q}e^{+iqd}+\frac{K}{N}\sum\limits_{q}U_qU_{-q}=\\
&=\frac{1}{2mN}\sum\limits_qP_qP_{-q}+\frac{2K}{N}\sum\limits_{q>-q}U_qU_{-q}(1-cos(qd)).
\end{array}
$$
Here $K=m\w^2/2$, and the standard formula for summing of the geometric progression of exponents was used, which gives zero for $q\neq q'$, and also we put pairs $q,-q$ into order so that explicitly written is only a half: for which $q>q'$- that gives coefficient 2 in the last summand. 

We pass again to the new variables, real numbers:

$$
\begin{array}{ll}
&U_q=X_q+iY_q,\ X_q=\frac{U_q+U_{-q}}{2},\ Y_q=\frac{U_q-U_{-q}}{2i};\\
&X_q=\frac{1}{\sqrt N}\sum\limits_nu_ncos(qnd),\ Y_q=\frac{1}{\sqrt N}\sum\limits_nu_nsin(qnd),\\
&\frac{\partial}{\partial U_q}=\frac{\partial}{\partial X_q}\frac{1}{2}+\frac{\partial}{\partial Y_q}\frac{1}{2i},\\
&\frac{\partial}{\partial U_{-q}}=\frac{\partial}{\partial X_q}\frac{1}{2}-\frac{\partial}{\partial Y_q}\frac{1}{2i},\\
&\frac{\partial^2}{\partial U_q\partial U_{-q}}=\frac{1}{4}\left(\frac{\partial^2}{\partial X_q^2}+\frac{\partial^2}{\partial Y_q^2}\right).
\end{array}
$$
A last obtain
$$
H=-\frac{1}{4mN}\sum\limits_{q>q'}\left(\frac{\partial^2}{\partial X_q^2}+\frac{\partial^2}{\partial Y_q^2}\right)+\frac{2K}{N}\sum\limits_{q>-q}(X_q^2+Y_q^2)(1-cos(qd)).
$$

We see that in the new coordinates our system represents the set of independent harmonic oscillators of the mass $\tilde m=2m$, with new coefficient $\tilde K=2K(1-cos(qd))$ and frequences  
\begin{equation}
\tilde\w_q=\sqrt{\frac{2K}{m}(1-cos(qd)}
\label{frequences}
\end{equation}

It means that the ground state of these new oscillators, which are quasi-particles, is the product of ground states of the form  
$$
\Psi_n(x)=\frac{1}{\sqrt{2^nn!}}\left(\frac{m\w}{\pi h}\right)^{1/4}exp\left( -\frac{m\w x^2}{2h}\right)H_n(x\sqrt{m\w /h})
$$
 ($x$- amplitude, $n=0$) of each of them separately (the same is true for excited states: they are excited independently of each other). Since Hermitian polynamials have the form  $H_0=1,H_1=x,H_2=x^2-1$, the ground state (with $H_0$) can be written as
$$
\Psi_0=\prod\limits_q\left(\frac{m\w_q}{\pi h}\right)^{1/2}exp\left(-\frac{m\w_q(X_q^2+Y_q^2)}{2h}\right) .
$$
Transforming this into old coordinates we have
\begin{equation}
\Psi_0=\prod\limits_q\left(\frac{m\w_q}{\pi h}\right)^{1/2}exp\left\{-\frac{m\w_q}{2hN}[(\sum\limits_nu_ncos(qnd))^2+(\sum\limits_nu_nsin(qnd))^2]\right\} .
\label{ground_state}
\end{equation}
The expression inside square brackets is $U_qU_{-q}$, and if all $\w_q$ be equal it would not be entangled state, because $\sum\limits_qU_qU_{-q}=\sum\limits_nu_n^2$. However, accordingly to (\ref{frequences}) all $\w_q$ are different and this reasoning is not right. The ground state of the system of dependent harmonic oscillators on the level of their initial coordinates turns entangled. 

We estimate the measure of entanglement by the formula (\ref{ent_measure}). It gives the integral 
$$
E_{i,j}=\int|\Psi_0|^2S_0|du_1du_2\ldots du_n
$$
where $S_0=\sum\limits_q-\frac{m\w_q}{hN}cos((i-j)qd)$. Since $S_0$ does not depend on $u$, we found that $E_{i,j}=S_0$. The ground state is entangled because $S_0>0$. For the first excited state with one phonon in $s$-th cosine mode:
$$
\begin{array}{ll}
\Psi_{1s}&=\prod\limits_q\left(\frac{m\w_q}{\pi h}\right)^{1/2}exp\left(-\frac{m\w_q}{2hN}\left( (\sum\limits_nu_ncos(qnd))^2+(\sum\limits_nu_nsin(qnd))^2\right)\right)\\
&\frac{1}{\sqrt{N}}(\sum\limits_nu_ncos(snd))
\end{array}
$$
Now the application of the definition (\ref{ent_measure}) after simple transformations gives
$$
E_{i,j}(\Psi_{1s})=\int ||\Psi_{1s}|^2S_0-|\Psi_0|^2cos(sid)cos(sjd)|du_1du_2\ldots du_n >0.
$$
Here we cannot complete without numerical computation; the best way is Monte Carlo method for many dimensional integrals. 

For the case of two excitations we have two possibilities: a) excitations of the different modes, or b) excitations of the same mode. For these cases we obtain correspondingly:
\begin{equation}
\begin{array}{ll}
E_{i,j}(\Psi_{2sp})&=\int ||\Psi_{2sp}|^2S_0-\frac{1}{\sqrt{N}}|\Psi_{1p}|^2cos(sid)cos(sjd)-\frac{1}{\sqrt{N}}\Psi_{1s}|^2cos(pid)cos(pjd)|\\
&du_1du_2\ldots du_n,\\
E_{i,j}(\Psi_{2s})&=\int ||\Psi_{2s}|^2S_0+|\Psi_0|^22u_icos(sid)2u_jcos(sjd)((\sum\limits_ncos(snd))^2-3)|
\\&du_1du_2\ldots du_n,
\end{array}
\end{equation}

Numerical estimations show that the measure of entanglement becomes more complicated with the growth of excitation.

\section{Entanglement in the combined systems}

With the purpose to find the adequate description of quantum state dynamics we must go beyond the limits of simple quantum mechanics, and include the electromagnetic field into consideration. 

Electromagnetic field is determined by its potential

$$
(\phi,A)
$$
here $\phi$ is the scalar field, $A$ is vector field. Electric and magnetic fields are expressed as follows:
$$
E=-\nabla\phi-\frac{\partial A}{\partial t},\ B=\nabla\times A.
$$
If for a given function $\psi$ we change vector and scalar field as
$$
A'=A+\nabla\psi,\ \phi '=\phi-\frac{\partial\psi}{\partial t}
$$ 
it will have no physical sense; this transformation is thus called calibration and we are free to choose it at our convenience. We further fix Lorenz calibration in order to write Maxwell equations briefly:
$$
\Box A_\mu=j_\mu, \ \nabla_\mu j_\mu=0,\ \nabla_\mu A_\mu=0.
$$
where $A_\mu =(\phi,A_x,A_y,A_z)$ and $\Box=\nabla_\mu\nabla_\mu=\frac{\partial^2}{\partial t^2}-\nabla^2.$

Quantization of the system ''field + particles'' presumes that a quantum state of this system has the form $\Psi (A_\mu (R,t),\bar q(t))$ where $A_\mu$ is the field, detremined in each point of space-time, $\bar q(t)$ are trajectories of all charged particles. Unitary law of evolution for $\Psi$ in non relativistic case (when $t$ we consider as the real time for all particles) can be represented in the form of Feynman path integrals:
$$
\Psi (A_{fin},\bar q_{fin})=\int K(A_{fin},\bar q_{fin},A_{in},\bar q_{in})\Psi (A_{in},\bar q_{in})DA_{fin}D\bar q_{in}
$$
where the kernel $K$ results from the integration along all trajectories $\g :\ A(t),\bar q(t),\ t\in [t_{in},t_{fin}]$, leading from the initial classical state $A_{in},\bar q_{in}$ to the final classical state $A_{fin},\bar q_{fin}$ of our 
system:
$$
K=\int exp(-\frac{i}{h}S[\g ])D\g
$$
where the action of ''field + particles'' has the form
$$
S[\g ]=\sum\limits_i\frac{m_i}{2}\int |\dot{q_i}|^2dt+\int\int [\rho (R,t)\phi (R,t)-\frac{1}{c}j(R,t)A(R,t)]d^3Rdt+\frac{1}{8\pi}\int (E^2-B^2)d^3Rdt 
$$
For electrodynamics the canonic transformation has the form:
\begin{equation}
\begin{array}{lll}
A(R,t)&=\sqrt{4\pi}c\int a_k(t)e^{ikR}\frac{d^3k}{(2\pi )^3},\ &a_k(t)=\frac{1}{\sqrt{4\pi}c}\int A(R,t)e^{-ikR}dR^3\\
\phi (R,t)&=\int \phi_k(t)e^{ikR}\frac{d^3k}{(2\pi )^3},\ 
&\phi_k(t)=\int \phi (R,t)e^{-ikR}\\
j(R,t)&=\int j_k(t)e^{ikR}\frac{d^3k}{(2\pi )^3},\ 
&j_k(t)=\int j(R,t)e^{-ikR}\\
\rho (R,t)&=\int\rho_k(t)e^{ikR}\frac{d^3k}{(2\pi )^3},\ 
&\rho_k(t)=\int\rho (R,t)e^{-ikR},
\end{array}
\label{can_field}
\end{equation}

Maxwell equations then can be rewritten in the form $\ddot{a}_{1k}+k^2c^2a_{1k}=\sqrt{4\pi}j_{1k}$ and the same for the second polarization, which means that the field is represented as the pairs of complex oscillators (see (\cite{FH}). 

What is the entanglement of the electromagnetic field? At first we consider the pure field without particles. Its psi-function we can represent as $\prod_k\Psi_{n_i}(a_k)$ where $\Psi_n$ is $n$-th excited level of harmonic oscillator. If we pass to the real amplitudes, $k$ and $-k$ values of impulse will be grouped and for this ''joint'' value of $k$ we have four independent real modes of oscillators, whereas is we distinguish $k$ and $-k$ we will have two independent modes, which we can distribute between two directions of polarisation: 1 and 2. Entanglement of the field we can estimate by the measure $E_{ij}$ if we applied it to $R$- representation of the field. For example, the entanglement of the ground state (without photons) is $E_{ij}(\Psi_0)$ where
$$
\Psi_0=\prod\limits_k\w_k^{1/2}exp\{ -\frac{m|kc|}{2hN}[(\sum\limits_Ra_rcos(kRd))^2+(\sum\limits_Ra_rsin(kRd))^2].
$$
The direct computation gives 
$$
E_{ij}(\Psi_0)=\sum\limits_k\frac{m|kc|}{2hN}cos[kd(R_j-R_i)]=0,
$$
because the sum is the real part of integral $\int exp(ikR_\Delta )dk$, for which integration over sphere $S^2:\ |k|=k_0$ gives zero for any $k_0$ for $R_\Delta =R_j- R_i\neq 0$.

On the other side, the analogous estimation for the field with only one photon (the single excited mode) shows non zero entanglement; the field with the single photon in each mode shows the entanglement in coordinate space between only pairs of the form $R,-R$. 

The entanglement in $R$- representation of the wave function has the direct connection with the entanglement of charged particles, interacting with the field. This entanglement shows the potential entanglement between the different charged particles, which can absorb or emit the photon. For example, let we be given three particles, each of which can absorb a photon and pass from the ground ($|0\rangle$) to the excited state ($|1\rangle$). We then obtain the state of the form 

\begin{equation}
\a (|100\rangle+|010\rangle+|100\rangle )+\b |000\rangle .
\label{single_photon}
\end{equation}
Of course, the degree of entanglement between particles will not be equal to $E_{ij}$, the last is only the indicator (''finger print'') of the ''real'' entanglement of the charged particles.

\section{Quasi-particle as the shield of entanglement}

Entanglement is the fundamental phenomenon, which was demonstrated in numerous experiments. Its deep nature follows from non-locality of quantum states, which entanglement is sufficienty large. Its first visible application concerns quantum computers, namely the realization of Grover algorithm (\cite{Gr}). For this aim we have to build quantum states of the form 
\begin{equation}
|\Psi\rangle=\a\sum\limits_{j\neq j_0}|j\rangle +\b |j_0\rangle
\label{Gro}
\end{equation}
where $\a^2(N-1)+\b^2=1$. The demonstration of such states would already mean that we can realize Grover search algorithm, which is such evolution of $|\Psi\rangle$ that $\b\ar\infty$. 

Can we realize this algorithm by means of only charged particles of oscillator type? 

Lagranjian of the oscillator with the external force $f$ has the form
$$
L=\frac{m}{2}\dot{u}^2-\frac{m\w^2}{2}u^2+f(t)u
$$

and its kernel with the interaction (see \cite{FH}) is:
\begin{equation}
K^{osc}=\sqrt{\frac{m\w}{2\pi h\ sin\ wT}}exp(\frac{i}{h}S^{osc}_{classic})
\label{kernelosc}
\end{equation}
where the classical action of the oscillator is
\begin{equation}
\begin{array}{ll}
S^{osc}_{classic}=&\frac{m\w}{2sin\ \w T}[ cos(\w T(u_f+u_i)-2u_fu_i\\
&+\frac{2u_f}{m\w}\int\limits^{t_f}_{t_i}f(t)sin\ \w(t-t_i)dt+\frac{2u_i}{m\w}\int\limits^{t_f}_{t_i}f(t)sin\ \w(t_f-t)d
t\\
&-\frac{2}{m^2\w^2}\int\limits_{t_i}^{t_f}\int\limits_{t_i}^tf(t)f(s)sin\ \w (t_f-t)sin\w (s-t_i)ds\ dt],\\
&\ \ \ \ T=t_f-t_i.
\end{array}
\label{classic_kernel}
\end{equation}

If the oscillator is the photon mode and external force comes from the charged particle $K^{osc}=K^A$ turns to the photon kernel for the fixed trajectory of particles and external force $f=j_{k\ 1(2)}$ of photon oscillator - to the current of charge.  All currents we must sum up ocer all participating particles.

We now consider the model with the conformation of chain of atoms. For example, it can be a protein of nuclear acid, for which the conformation - the spatial location of the chain - is conserved by hydroden bonds between segments of the chain. Let the conformation be given by the three dimension vector function of the real parameter $\tau$ of the form $q=q(\tau )$. We treat that oscillators (electrons) are located along this chain so that their small shifts do not depend on the conformation of the whole chain in the space. Electrons are in quantum states as well, but we account only the oscillations of their density. The parameter $\tau$ thus plays the role of $n$ - number of oscillator from the model example (see the previous section). We denote by $u_\tau$ the shift of the oscillator corresponding to the value $\tau$. The density of charge (electronic cloud) we denote by $\rho (q)$. It velocity is $\dot{u_{\tau}}$.

If the charge $e$ was point wise and was located in the point $q$, its density $\rho (R)$ would be delta-function $\rho=e\delta (R-q)$, and the current created by its movement would be equal $e\dot{q}\delta (R-q)$. Since Fourier transform of delta-function equals $e^{-ikq}$, we conclude that in the case of one point wise change, which performs small oscillations arond the point $q$
\begin{equation}
U_k=e(q)u(q)e^{-ikq},\ j_k=e(q)\dot{U_k}=e(q)u(q)e^{-ikq}
\label{dotcharge}
\end{equation}
Canonic transform for the system of oscillators looks as follows:
$$
U_k=\sum\limits_nu_ne^{-ikn}
$$
and the velocity 
$$
\dot{U_k}=\sum\limits_n\dot{u}_ne^{-ikn}.
$$
From (\ref{dotcharge}) we obtain for the oscillations of the continuous distribution of electronic density
\begin{equation}
\begin{array}{ll}
&U_k=\int\limits_\tau \rho (q(\tau)) u_\tau e^{-ikq(\tau )}d\tau\\
&j_k=\int\limits_\tau\rho (q(\tau ))\dot{u_\tau}e^{-ikq(\tau )}d\tau
\end{array}
\label{current}
\end{equation}
where the dot as usually, means the differentiation on the time. The coordinate  $q$ depends on the internal coordinate on the chain $\tau$, but for the fixed conformation of the chain in space $q$ doe not depend on the time $t$. If we treat the deformation of the chain, the dependence $q_\tau (t)$ from the time arises. However, its dynamics is much more slow, than the dependence $u_\tau (t)$. The dependence $u$ from the time determins phonons in the chain, and at last, the emission of photons that is much more rapid process than the change of conformation $q_\tau$, hence at the work with quasi-particles we ignore the dynamics of conformation. 

We now find the Lagranjian of the system: ''3-dimensional particle + field'', following the analogy with the already considered examples. The current $j_k=$ $F_k/m_{phot}\sqrt{4\pi},$  $k=|k|=k_{phot}$, where $m_{phot}$ is the fictive ''mass of the photon''. From the other side $j_k=e\dot{U}_k$ where $U_k$ is the coordinate in $k$- basis of the sfift of particle-oscillator of the mass $\tilde m=m_{part}$ and the coefficient of elasticity $K=K_{part}$ respectively to its equilibrium position. 

The general Lagranjian $L_k=L_{1k}$ or $L_{2k}$- in dependency of the polarisation of the field, which we omit, is
The \begin{equation}
\begin{array}{ll}
L_k=&\frac{m_{phot}\dot{a}_k}{2}-\frac{m_{phot}k^2c^2}{2}+\sqrt{4\pi}ea_k\dot{U}_km_{phot}\\
&\frac{m_{part}}{2}\dot{U}_k^2-\frac{m_{part}\w_{k\ part}^2}{2}U^2_k
\end{array}
\label{preliminary}
\end{equation}
We note that $m_{phot}$ gives only the multiplier in the kernel and thus can be ignored.

By the analogy of one particle Lagranjian (\ref{preliminary}) we can compose Lagranjian for the chain with the field and write the action and the kernel analogously to formulas (\ref{kernelosc}), taking into account already found action of oscillators of the field at the movement of them along the classical trajectory and the fixed movement of particles:
(\ref{classic_kernel}). Here for the currents and shifts of electronic density the expression (\ref{current}) should be used.

The result looks as follows. The kernel of the system ''chain + field'' has the form:

\begin{equation}
\begin{array}{ll}
&K^k_{gen}(t)=\int\limits_{\g_{part}}exp(\frac{i}{h}\int\limits_{t_i}^t\frac{m_{part}}{2}(j_k^2-\w^2_{k\ part}U_k^2)K^k_{phot}(t)\ DU_k,\\
&K_{phot}^k(t)=\sqrt{\frac{m_{phot}rc}{2\pi ih\ sin(kct)}}exp(\frac{i}{h}S_{phot}^{class}[U_k(\tau )]),\\
&S_{phot}^{class}[U_k(\tau)]=\frac{m_{phot}kc}{2\ sin(kct)}[ cos(kct)(a_{k\ fin}^2+a_{k\ in}^2)\\
&-a_{k\ fin}a_{k\ in}+2\frac{a_{k\ fin}}{m_{phot}kc}\int\limits_{t_{in}}^t m_{phot}\sqrt{4\pi}j_k(b)sin(kc(b-t_{in}))db\\
&+2\frac{a_{k\ in}}{m_{phot}kc}\int\limits_{t_{in}}^t m_{phot}\sqrt{4\pi}j_k(b)sin(kc(b-t_{in}))db\\
&-\frac{2}{m_{phot}^2k^2c^2}\int\limits_{t_{in}}^t\int\limits_{t_in}^bj_k(b)
j_k(s)m_{phot}^24\pi sin(kc(t-b))sin(kc(s-t_0)ds\ db]
\end{array}
\label{ker}
\end{equation}

We see that there is the complete parallelism of the computation on the pairs of the form $k,-k$, which does not take place in the general case. Peculiarity of our case is that particles are included into threads that makes possible the complete passage into  $k$- representation for the field and for particles as well. Quantum dynamics of the peculiar case we have just considered can be thus effectively simulated on classical computers. On the other hand, the ability to produce of the states (\ref{Gro}) means the ability to fulfil Grover search algorithm, which cannot be effectively simulated on classical computers (see \cite{BBBV},\cite{Oz}). It proves, that to produce the states of the form (\ref{Gro}) via quantum processors of the type ''chains + field'' is impossible. 

Entanglement of oscillator type is shown in experiments with ions in trap (see \cite{Zo}); this is the states of W-type: $|100\rangle+|010\rangle+|001\rangle$ and GHZ-type $|000\rangle+|111\rangle$. The oscillator character of these types of entangled states follows from the possibility to obtain them by wave-like process. For example, the single photon can cause excitation of only one of ions (W-type); for the example from above (\ref{single_photon}) it is possible to make $\b$ negligible by using resonators. These types is entanglement, protected into quasi-particles of a system of oscillators. We have shown that purely oscillator type quantum processor is not sufficient to build the robust quantum computer. 

It is important to obtain the states of the form (\ref{Gro}), because they do not belong to the oscillator type. The experimental demonstration of such states would be the decisive argument for the practical validity of the direct approach to quantum computer - via quantum gates (see \cite{Fe}).

\section{Decoherence and energy gap}

The source of decoherence is the entanglement between the considered system and its environment caused by photon emission. In the abstract quantum computing the formal way to suppress decoherence is quantum error correcting codes (\cite{Sh} and many others) . For this it is necessary to have the robust quantum devices with few tens qubits that requires resistence against decoherence on the physical level. We now estimate the degree of decoherence via entanglement with the emitted photon of the simgle atom and show that the dependence on its frequency is approximately linear. It shows that we can control decoherence by avioding of passages with big energy gaps even if real phonots are emitted by a quantum device.

Let $\Psi_i$ and $\Psi_f$ be the initial and final states of the atom correspondingly, $\w_{if}=(E_i-E_f)/h$ be the frequency of the passage. Let the polarization of the emitted photon be denoted by $\bar e$ , and the only restiction on its impulse direction is that is orthogonal to $\bar e$. The final state of the atom and electromagnetic field will be 
$$
\Psi_{general}=\sum\limits_{k:\ |kc|=\w_{if}}\la_k\Psi_f\Psi_{phot\ k}
$$
Here the phase of $\la_k$ does not influence to the entanglement if we use (\ref{canonic_ent}).
For the small frequences ($\w_{if}^{-1}$ larger than the atom size) we can use the dipole approximation: $p_k=|\la_{if}|^2\approx \w_{if}(\bar e\mu_{if} )^2$, where $\mu_{if}$ is the dipole electric momentum of the atom (\cite{FH}).
The entanglement measure between the atom and the electromagetic field is thus the integral on the two dimension sphere 
$$
{\cal M}_{if}=-\int\limits_{S^2} 4\pi^2\w_{if}(\bar e\bar \mu )^2dS_{\bar e}
$$
where $S_{\bar e}$ is the element of square on the surface of $S^2$ corresponding to $\bar e$. We straightforwardly obtain that 
$$
{\cal M})_{if}=-4\pi^2\w_{if}[A\ ln(4\pi^2\w_{if})+B]
$$
where $A=\int_{S^2}(\bar e\bar \mu )^2dS_{\bar e}$, $B=\int_{S^2}2(\bar e\bar \mu )^2ln(|\bar e\bar \mu |)dS_{\bar e}$. The second root of the quadratic polynamial has no physical sense because it does not belong to the area of validity of the dipole approximation. It shows that the dependence of entanglement (and the decoherence resulted from it) is proportional to the energy gap corresponding to the considered passage.

\section{Conclusion}

We have evaluated the bi-partite entanglement in a system of interacting harmonic oscillators, for which all eigenstates beginning from the ground state are entangled, and have shown that the entanglement of spatial components of electromagnetic field can serve as a ''finger print'' of the entanglement between real particles, which emit the field. Entanglement can be estimated for systems of particles and electromagnetic field where it turns linear from the energy gap.
We have proved that the system of oscillators with the field cannot play the role of quantum computer, because its evolution can be effectively simulated on a classical computer. The entangled states obtained in experiments have the type of oscillator states, whereas Grover search algorithm requires states of the other type. Grover states cannot arise in a system based only on harmonic oscillators, and the experimental demonstration of them represents the significant step on the way to quantum computer.

\end{document}